\documentclass{aa} 

\DeclareRobustCommand{\rchi}{{\mathpalette\irchi\relax}}
\newcommand{\irchi}[2]{\raisebox{\depth}{$#1\chi$}} 

\newcommand{\rlight}{r_\textrm{L}}

\usepackage{amsmath}
\usepackage{graphicx}
\usepackage{txfonts}
\usepackage{hyperref}
\usepackage{color}

\begin{document} 

\title{A double dipole geometry for PSR~J0740+6620}


\author{J. P\'etri\inst{1}
         \and S. Guillot \inst{2}
         \and L. Guillemot \inst{3,4}
         \and D. Gonz\'alez-Caniulef \inst{2}
         \and F. Jankowski \inst{3}
        \and J.-M. Grie{\ss}meier \inst{3,4}
         \and G. Theureau \inst{3,4}
         \and I.~Cognard\inst{3,4}
          }

   \institute{Universit\'e de Strasbourg, CNRS, Observatoire astronomique de Strasbourg, UMR 7550, F-67000 Strasbourg, France.\\
	\email{jerome.petri@astro.unistra.fr}   
	\and 
	IRAP, CNRS, Universit\'{e} de Toulouse, CNES, 9 avenue du Colonel Roche, BP 44346, F-31028 Toulouse Cedex 4, France
	\and
	LPC2E, OSUC, Univ Orl\'eans, CNRS, CNES, Observatoire de Paris, F-45071 Orl\'eans, France
	\and 
	ORN, Observatoire de Paris, Universit\'e PSL, Univ Orl\'eans, CNRS, 18330 Nan\c{c}ay, France
}

   \date{Received ; accepted }

 
  \abstract
   {Millisecond pulsars are known to show complex radio pulse profiles and polarisation position angle evolution with rotational phase. Small scale surface magnetic fields and multipolar components are believed to be responsible for this complexity due to the radiation mechanisms occurring close to the stellar surface but within the relatively small light-cylinder compared to the stellar radius.}
   {In this work, we use the latest NICER phase aligned thermal X-ray pulse profile of PSR~J0740+6620 combined with radio and $\gamma$-ray pulse profiles and radio polarisation to deduce the best magnetic field configuration that can simultaneously reproduce the light-curves in these respective bands.}
   {We assume a polar cap model for the radio emission and use the rotating vector model for the associated polarisation, a striped wind model for the $\gamma$-ray light-curves and rely on the NICER collaboration results for the hot spot geometry.}
   {We demonstrate that an almost centred dipole can account for the hot spot location with a magnetic obliquity of $\alpha \approx 51 \degr$ and a line of sight inclination angle of $\zeta \approx 82 \degr$. However, with this geometry, the hot spot areas are three times too large. We found a better solution consisting of two dipoles located just below the surface in approximately antipodal positions.}
   {Our double dipole model is able to reproduce all the salient radio and $\gamma$-ray characteristics of PSR~J0740+6620 including radio polarisation data. A double dipole solution is more flexible than an off-centred dipole because of two independent magnetic axes and could hint at a magnetic field mostly concentrated within the crust and not in the core.}

\keywords{Magnetic fields -- Methods: numerical -- Stars: general -- Stars: rotation  }

\maketitle

%

\section{Introduction}

Thanks to multi-wavelength observations of pulsar emission, a global picture of the neutron star magnetosphere and its radiation mechanisms is emerging. Whereas for non-recycled pulsars, the radio emission is well constrained to occur along dipolar magnetic field lines at several tens of stellar radii \citep{johnston_thousand-pulsar-array_2023}, the $\gamma$-ray emission appears to emanate from the striped wind just beyond the light-cylinder \citep{petri_young_2021} and the non-thermal X-ray emission between those two regions \citep{petri_localizing_2024}. However, the case of millisecond pulsars (MSPs) still resists to this interpretation mainly because of the small size of their magnetosphere, typically of the order of several light-cylinder radii, and the possible influence of multipolar components at low altitude. The radio pulse profiles of MSPs have large widths and show a complex structure as opposed to young pulsars. Moreover their radio polarisation position angle (PPA) evolution with phase usually does not follow the rotating vector model (RVM) \citep{radhakrishnan_magnetic_1969} although the $\gamma$-ray pulse profiles are undistinguishable from normal pulsars, allowing to fit them in a similar manner \citep{benli_constraining_2021}. Some hints about the presence of non dipolar magnetic field components close to the surface are inferred from the hot spot geometry deduced by the thermal X-ray emission. The Neutron star Interior Composition ExploreR (NICER) collaboration has already investigated four MSPs and deduced their hot spot shape \citep{riley_nicer_2019, salmi_nicer_2024, salmi_radius_2022, choudhury_nicer_2024, miller_psr_2019, miller_radius_2021, dittmann_more_2024, vinciguerra_updated_2024} showing some of them with non-axisymmetric geometries. 

In this paper we focus on PSR~J0740+6620 (with its X-ray pulsation discovered by \citealt{wolff_nicer_2021}), the second MSP with pulse-profile modelling analyses published by the NICER team \citep{miller_radius_2021, riley_nicer_2021} and reinvestigated recently by \cite{salmi_radius_2024, dittmann_more_2024}. PSR~J0740+6620 is also a $\gamma$-ray pulsar \citep{guillemot_gamma-ray_2016}, evolving in a binary system with good constraints on the orbital parameters \citep{cromartie_relativistic_2020,fonseca_refined_2021}. Among them the Shapiro delay has been measured and the orbital inclination angle~$i$ deduced to be $i\approx 87.4\degr$. 
The \textit{Fermi} Large Area Telescope (LAT) Third Catalog of $\gamma$-ray pulsars \citep[3PC,][]{smith_third_2023} contains a wealth of information, notably light-curves and spectra, useful to understand the high-energy processes, particle acceleration and radiation within the neutron star magnetosphere and wind. Accurate $\gamma$-ray light-curves are available for PSR~J0740+6620, making it a good target for our purpose to model the magnetic field structure from the surface to the light-cylinder and beyond. Moreover it radiates also in radio and its PPA has been measured by the Green Bank Telescope \citep{wahl_nanograv_2022} and by the Nan\c{c}ay Radio Telescope (NRT), as shown in the paper. Thus, we follow the same line of reasoning as \cite{petri_constraining_2023-2} who studied the magnetic field topology of the first NICER pulsar PSR~J0030+0451. In this work, we investigate the magnetic field configuration of PSR~J0740+6620, relying on the latest simulations detailed in \cite{petri_multi-wavelength_2024-3}. Section~\ref{sec:Observations} details the multi-wavelength data set used for this study to determine the magnetic field structure as exposed in Sec.~\ref{sec:magnetic_field}. A discussion of our results is presented in Sec.~\ref{sec:discussion} before concluding in Sec.~\ref{sec:conclusion}.

\section{Observations and data\label{sec:Observations}}

PSR~J0740+6620 is a MSP with spin period $P=2.89$~ms in an almost circular orbit of period $4.77$~day. Its orbital inclination angle, derived from Shapiro delay is $i\approx 87.4\degr$ \citep{cromartie_relativistic_2020}. PSR~J0740+6620 shows pulsation in a broad-band electromagnetic spectrum from the radio wavelength, through X-ray up to $\gamma$-ray. We briefly summarise the data used for this study. 

\subsection{NRT radio profile and \textit{Fermi} LAT $\gamma$-ray profile}

\begin{figure}[h]
	\centering
	\includegraphics[width=\linewidth]{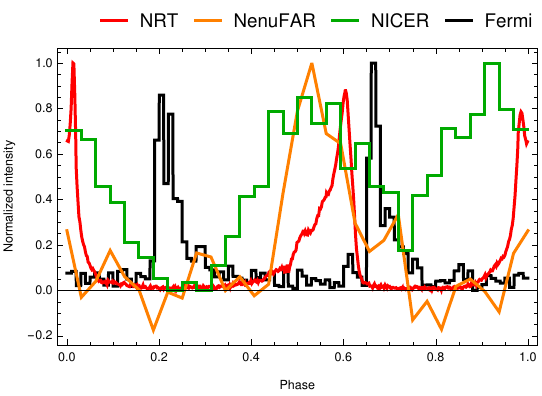}
	\caption{Multi-wavelength pulse profiles of PSR~J0740+6620, as observed in radio with NenuFAR (39 to 76 MHz, orange line) and the Nan\c{c}ay Radio Telescope (1.4~GHz, red line), in X-rays with NICER (energy band $0.3-1.5$~keV, green line) and in $\gamma$ rays with the \textit{Fermi} LAT ($\geq 0.1$ GeV, black line).}
	\label{fig:j0740+6620_multilambda}
\end{figure}

As mentioned in the introduction, PSR~J0740+6620 was reported to emit pulsed $\gamma$-ray emission by \citet{guillemot_gamma-ray_2016} and the pulsar was also included in the 3PC catalogue of \textit{Fermi} LAT pulsars \citep{smith_third_2023}. In Fig.~\ref{fig:j0740+6620_multilambda} we display the $1.4$~GHz NRT radio pulse profile and $E \geq 0.1$~GeV $\gamma$-ray pulse profile from 3PC, as taken from the 3PC auxiliary files archive\footnote{See \url{https://fermi.gsfc.nasa.gov/ssc/data/access/lat/3rd_PSR_catalog/} .}. We note that the NRT timing solution used for producing the \textit{Fermi} LAT pulse profile shown in Fig.~\ref{fig:j0740+6620_multilambda} is the same as that used for folding the NICER data, see Sect.~\ref{subsec:NICER}. Since the X-ray and the $\gamma$-ray data were folded with the same timing solution, the same radio fiducial phase was used for all analyses, guaranteeing that the relative phasing between the radio, X-ray and $\gamma$-ray components are indeed correct. This phase alignment is crucial for our light-curve fitting procedure.

\subsection{NenuFAR profile}

On 2024-03-13, we observed PSR~J0740+6620 with NenuFAR \cite[New extension in Nançay upgrading LOFAR, see][]{zarka_nenufar_2015}, Zarka et al. in prep., using the LUPPI backend \citep{bondonneau_pulsars_2021}. We recorded data from 39 to 76~MHz. We refolded the NenuFAR observations using the NRT timing solution, except for the DM value, for which the value determined directly from the NenuFAR observations is considerably more precise owing to the low-frequency coverage of NenuFAR.

PSR~J0740+6620 is among the twelve MSPs detected by NenuFAR, \citep{bondonneau_pulsars_2021}. The NenuFAR profile of the pulsar considerably differs from that obtained by NRT. It has only one peak (rather than two). The main peak seems to be wider at NenuFAR frequencies. At NRT frequencies, the main peak is clearly asymmetric, with a gradual leading edge and a step trailing edge. At NenuFAR frequencies, the low S/N does not allow definitive conclusions, but the mean peak seems to have a stepper leading edge and a gradual trailing edge. With the available data, we cannot rule out profile frequency evolution, making the comparison between the phases of the main peaks of NRT and NenuFAR difficult. Within error margins, our observations are compatible with the radio emission detected by NenuFAR and NRT being emitted at the same rotational phase. Also, the
low S/N of the NenuFAR observation did not allow for an analysis of polarisation properties at those frequencies.

\subsection{NICER pulse profile}
\label{subsec:NICER}

To produce the X-ray pulse profile from NICER data, we processed all data from 2018-09-21 to 2024-08-31 (ObsID 1031020101 to 7031020525), using the task \texttt{nicerl2} (from \texttt{NICERDAS v13.0} distributed with \texttt{HEASOFT v6.34}) and its default filtering criteria. The most recent calibration file (v20240206) was used for this step. Then, the package \texttt{NICERsoft} \footnote{\url{https://github.com/paulray/NICERsoft}} was used to include additional filtering conditions. Specifically, we excluded observing periods with 
1) Earth magnetic cut-off rigidity \texttt{COR\_SAX} $<1.5$~GeV/c, 
2) space weather index KP $\geq5$, 
3) overshoot rates larger than $1.5$~c/s, 
4) median undershoot rates larger than $100$~c/s.  Finally, a count-rate cut was performed to exclude all time intervals where the rate was larger than $1$~c/s in the $2-10$~keV range, where the pulsar thermal emission is negligible, which effectively removes all remaining periods of high background.  The pulse phase of each individual photon was then calculated with the task \texttt{photonphase} (of the package \texttt{PINT version v1.0} \footnote{\url{https://github.com/nanograv/PINT}}), using the timing solution from the 3PC catalogue of \textit{Fermi} LAT pulsars \citep{smith_third_2023}.  Finally, the event files of each ObsID were merged into a single event file to make the pulse profile of this pulsar with 32 phase bins (see Fig.~\ref{fig:j0740+6620_multilambda}).  The pulse profile was produced with 32 bins in the energy range $0.3-1.5$~keV, the energy range where the thermal emission is significant \citep{riley_nicer_2021, miller_radius_2021}.

\section{Magnetic field determination\label{sec:magnetic_field}}

The geometry of the dipole magnetic field of several MSPs has already been investigated with a simple striped wind model relying on the split monopole \citep{petri_unified_2011}. More recently \cite{benli_constraining_2021} used the numerical solution of the pulsar force-free magnetosphere to deduce the large scale dipole geometry for another sample of MSPs. The simplest deviation from a dipolar magnetosphere assumes an off-centred dipole, displaced with respect to the centre of the star. This configuration has been extensively studied analytically by \cite{petri_off-centred_2021} and \cite{petri_impact_2019}, including polarisation \citep{petri_polarized_2017}. However, as a simple first guess outside the light-cylinder, a centred dipole offers a minimalistic model with the fewest parameters to be fitted. We start with this assumption.

\subsection{Dipole geometry from $\gamma$-ray modelling}

\begin{figure}[h]
	\centering
	\includegraphics[width=0.9\linewidth]{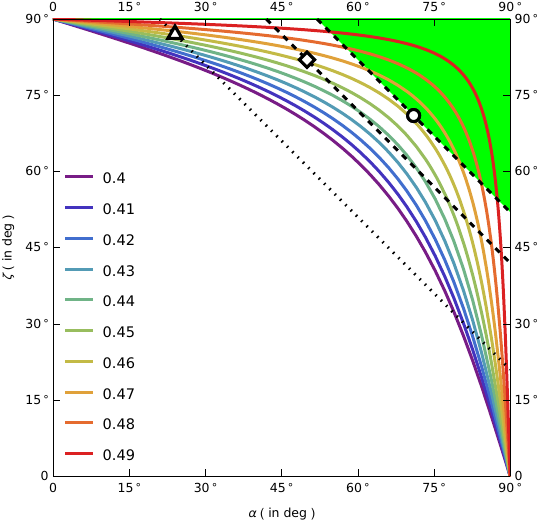}
	\caption{Constraints on the viewing angle and obliquity. For symmetry reasons, the other three quadrants are not shown. Colours indicate the value of the peak separation~$\Delta$ from $0.41$ in violet to $0.49$ in red. The white circle, square and triangle show the location of three possible fits to the $\gamma$-ray peak separation with $(\alpha,\zeta)$ respectively as $(71\degr,71\degr)$, $(51\degr,82\degr)$ and $(24\degr,87\degr)$. The green area highlight the region in the diagram where $\gamma$-ray emission is detected simultaneously with the radio pulse for $(\alpha,\zeta)=(71\degr,71\degr)$. The other dashed and dotted line delimit the same region but for $(\alpha,\zeta)=(51\degr,82\degr)$ and $(\alpha,\zeta)=(24\degr,87\degr)$ respectively.}
	\label{fig:obliqinclin}
\end{figure}

The $\gamma$-ray light-curve fitting procedure has been explained in depth in \cite{benli_constraining_2021} and \cite{petri_young_2021}. In summary, the two main parameters to be adjusted are the $\gamma$-ray light-curve peak separation~$\Delta$ and the phase lag~$\delta$ between the first $\gamma$-ray peak and the radio profile. The third \textit{Fermi} LAT pulsar catalogue reports a peak separation of $\Delta = 0.462$ and a radio time lag of $\delta = 0.196$. To a very good approximation, \cite{petri_unified_2011} showed that the peak separation~$\Delta$ is related to the pulsar obliquity $\alpha$ and line of sight inclination angle $\zeta$ by
\begin{equation}\label{eq:cos_pi_delta}
\cos (\pi \, \Delta) = |\cot \alpha \cot \zeta| \ .
\end{equation}
However, as in radio, the $\gamma$-ray profile may be asymmetric such that the phase location of the peak intensity does not necessarily correspond to the middle of the pulse profile, thus the measurement of $\Delta$ could be altered if another definition is used for the separation~$\Delta$. In the particular case of PSR~J0740+6620, we checked that the peak separation approach gives the same results as for the separation obtained from measuring the width at 30\% peak intensity except for a slight shift in phase, see Table~\ref{tab:phase_lag}. However the width at 20\% peak intensity leads to a smaller peak separation of $\Delta=0.44$. Moreover, as the radio pulse profile shows a pulse and an interpulse, this pulsar should be close to an orthogonal rotator. 
Let us however emphasise that a pulse with an interpulse could also be explained by an almost aligned rotator as demonstrated by \cite{hankins_interpulse_1981}. However because $\gamma$-rays are detected simultaneously with radio photons, such geometry is forbidden and the constraints on the line of sight $\zeta$ and obliquity $\alpha$ are more stringent, see the green shaded area in Fig.\ref{fig:obliqinclin}, and also figure~2 of \cite{petri_multi-wavelength_2024-3} for the condition to detect simultaneously radio and $\gamma$-rays. Moreover NICER pulse profile modelling gives a line of sight inclination angle of $\zeta \approx 87\degr$, assuming roughly spin-orbit alignment, which is also incompatible with an almost aligned rotator that we thus discarded.

Therefore the presence of a pulse and an interpulse puts stringent constraints on the geometry because the radio pulses emanating from both poles can only be detected if $|\zeta-\pi/2|<\rho$ and $|\alpha-\pi/2|<\rho$ are simultaneously satisfied where $\rho$ is the radio beam cone opening angle. This condition is given by the green shaded area in Fig.\ref{fig:obliqinclin} where we assume a radio beam cone half-opening angle of $\rho = 38 \degr$, the minimal opening angle required for detecting the main radio pulse and the interpulse if the $\gamma$ peak separation is $\Delta \approx 0.46$. We conclude that the beam size must be at least as large as $38 \degr$, because otherwise the peak separation would always be larger than~$\Delta \gtrsim 0.46$. 
The minimal cone opening angle is given by the geometry imposing $\zeta = \alpha$. For this particular configuration, setting $\alpha \approx \zeta$, we get $\cot \zeta \approx \cot \alpha \approx \sqrt{\cos (\pi \, \Delta)} \approx 0.344$. This leads to a first guess of the geometry given by $\alpha \approx \zeta \approx 71 \degr$. In order to get $\Delta \lesssim 0.462$, however, we need to impose a radio cone beam with half opening angle $\rho \gtrsim 38 \degr$ if either $\alpha$ or $\zeta$ is changed. Indeed, the NICER collaboration reports $\zeta \approx 87\degr$ implying larger radio cone beams. Taking $\zeta \approx 87\degr$ leads to an obliquity $\alpha \approx 24\degr$ and a cone opening angle of $\rho \approx 69\degr$. However, this geometry is incompatible with the detection of a main pulse and an interpulse. In any case, requiring a higher inclination of the line of sight would led to lower obliquities. For instance if $\zeta \approx 80 \degr$, we would require $\alpha\approx 56\degr$ and getting two radio pulses becomes less and less probable.

\begin{figure}[h]
	\centering
	\includegraphics[width=\linewidth]{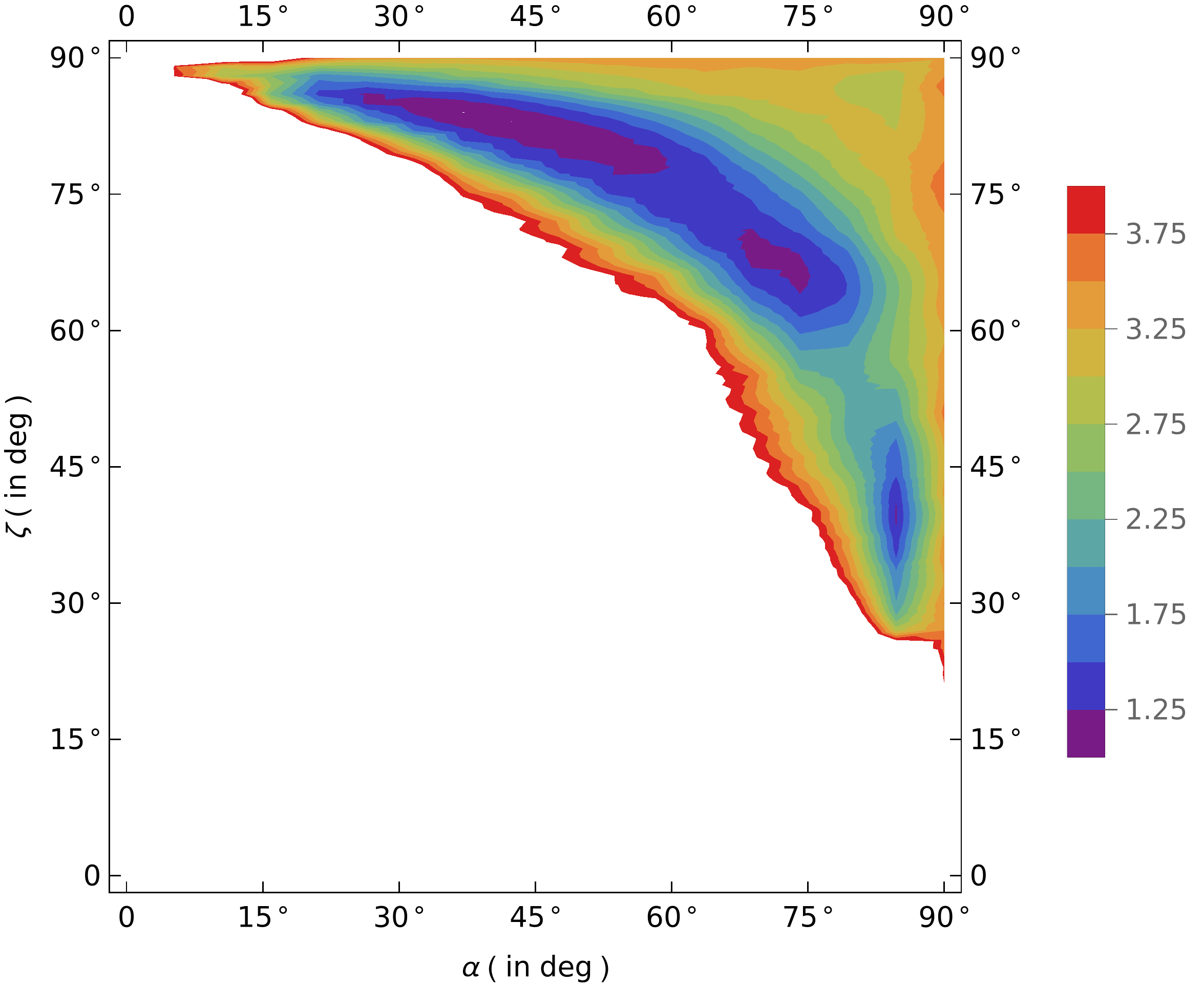}
	\caption{Best fit angles $\alpha$ and $\zeta$ for PSR~J0740+6620 $\gamma$-ray light-curve. The lowest values of $\rchi^2$, which is colour code in the legend, represents the preferred values.}
	\label{fig:gamma}
\end{figure}

\begin{figure}[h]
	\centering
	\includegraphics[width=\linewidth]{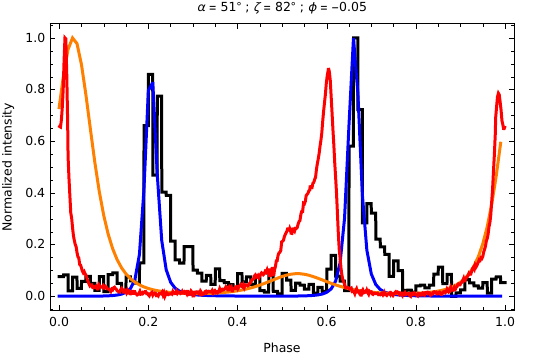}
	\caption{Example of good fit of the $\gamma$-ray pulse profile ($\geq$ 0.1 GeV). The radio pulse profile is shown in red, our model for the radio component is displayed in orange, the $\gamma$-ray profile is shown in black and the fit of the $\gamma$-ray component is shown in blue.}
	\label{fig:fitj00300451_2}
\end{figure}

In order to get independent constraints on these angles, we performed a fit of the $\gamma$-ray light-curve as shown in Fig.~\ref{fig:gamma}, using the same model as by \cite{benli_constraining_2021}. A good fit to this $\gamma$-ray light-curve with two radio pulses is given by $\alpha=51\degr$ and $\zeta=82\degr$, as shown in Fig.~\ref{fig:fitj00300451_2}. In order to accurately fit the $\gamma$-ray peak time lag with respect to the radio peak, we added a third parameter depicted by the phase shift and denoted by~$\phi$ as done in our previous works. Its value remains however small, $\phi=0.02$. The associated radio beam opening angle is $\rho\approx 50 \degr$, compatible with the detection of two radio pulses.
\begin{table}[h]
	\centering
	\caption{Phase locations of the first and second pulse profile peaks in radio, X-ray and $\gamma$-ray. \label{tab:phase_lag}}
	\begin{tabular}{llll}
		\hline
		Wavelength & Pulse~1 & Pulse~2 & $\Delta \phi$ \\ 
		\hline
		$\gamma$-ray peak & 0.19 -- 0.20 & 0.66 -- 0.67 & 0.46  \\
		$\gamma$-ray centre 30\% & 0.215 & 0.675 & 0.46 \\
		$\gamma$-ray centre 20\% & 0.24 & 0.68 & 0.44 \\
		X-ray peak       & 0.5 -- 0.53 & 0.90 -- 0.93 & 0.40  \\
		X-ray centre 30\% & 0.53 & 0.93 & 0.40 \\
		NRT centre 10\%     & -0.012 & 0.543 & 0.53  \\
		NRT centre 5\%  & -0.017 & 0.523 & 0.51  \\
		$[\phi_1; \phi_2]$  $(w_{10})$			 & [-0.073; 0.050] & [0.454;0.632] &  \\
		&  0.123 & 0.178 & \\
		$[\phi_1; \phi_2]$  $(w_{5})$			 & [-0.099; 0.065] & [0.407;0.640] &  \\
		&  0.164 & 0.233 & \\
		\hline
	\end{tabular}
	\tablefoot{The phase difference~$\Delta \phi$ between both peaks is also shown (restricted to the interval [0,0.5]). The $w_{10}$ line indicates the phase interval where the radio pulses are detected above 10\% of maximal intensity and similarly $w_5$ at 5\%.}
\end{table}

On the one hand, radio and $\gamma$-rays furnish some constraints on the geometry. On the other hand, X-ray pulse profile modelling is nowadays able to deduce the polar cap shapes, locations and temperatures detected as atmospheric emission from hot spots. Actually, these hot spots are heated by the bombardment of charged particles produced in the magnetosphere and partly returning to the stellar surface. These particles are also responsible for the radio emission. There is therefore a one to one correspondence between polar caps and hot spots. These measurements by the NICER collaboration give independent constraints on the dipole or multipole magnetic field geometry, which we exploit in the following. However, we stress that there is a degeneracy in the geometry found for the two hot spots because they can be flipped.

\subsection{Dipole geometry from X-ray pulsations}

For the MSP PSR~J0740+6620, \cite{riley_nicer_2021} report a radius $R=12.39$~km and a mass $M=2.07\ M_\odot$, thus a ratio of stellar radius to light-cylinder radius ${ a = R/\rlight \approx 0.091}$ with $\rlight=c/\Omega$. Based on the pulse profile evolution with energy, they also deduced the polar cap geometry assuming only a pair of circular hot spots. Moreover, the relatively low signal to noise ratio of the data did not allow testing more complex hot spot shapes. Independent constraints from a second group \citep{miller_radius_2021} were also reported. We use the latest results of \cite{salmi_radius_2024} and \cite{dittmann_more_2024} as best fit parameters, summarised in Table~\ref{tab:NICER1} as the median of the posterior distributions ('median') and the value at maximum likelihood (ML), which may differ from the median for non-symmetric distributions. The parameters $(\Theta_{\rm p},\Theta_{\rm s})$ on one side and $(\phi_{\rm p}, \phi_{\rm s})$ on the other side define the centre of the primary ${\rm p}$ and secondary ${\rm s}$ hot spot colatitude and initial phase whereas $\zeta_{\rm p}$ and $\zeta_{\rm s}$ are their angular sizes. The geometry of the off-centred dipole with the parameters $(\alpha, \beta, \delta, \epsilon)$ defined in \cite{petri_polarized_2017} is then deduced and shown on the right columns in the same table. We found a mildly off-centred dipole structure with a displacement $d$ such that $\epsilon=d/R\approx0.22$.

The magnetic obliquity is $\alpha\approx 72\degr$ in the median case and $\alpha\approx 78\degr$ in the ML case for the \cite{salmi_radius_2024} results and a similar geometry for the ML case by \cite{dittmann_more_2024}. This is however $20\degr$ more than the obliquity found by $\gamma$-ray light-curve fitting, although the inclination angle is within $5\degr$ of the NICER results. It would be enlightening to redo the X-ray pulse profile modelling and observe the variation induced on $\alpha$ by removing the constraint on $\zeta$, allowing for lower values for instance closer to $\zeta=80\degr$ and to check whether these new values would be more compatible with the previous $\gamma$-ray fit.

\begin{table*}[h]
	\centering
\caption{Off-centred dipole geometry deduced from the polar cap location after the new joint NICER and XMM-Newton results of [1] \citep{salmi_radius_2024} and from [2] \citep{dittmann_more_2024}. \label{tab:NICER1}}
\begin{tabular}{lccccccccccccl}
	\hline
	Hot spot & $\Theta_{\rm p}$ & $\phi_{\rm p}$ & $\xi_{\rm p}$ & $\Theta_{\rm s}$ & $\phi_{\rm s}$ & $\xi_{\rm s}$ & $\xi_{\rm s}/\xi_{\rm p}$ & $\zeta$ & $\alpha$ & $\beta$ & $\delta$ & $\epsilon$ & Ref \\
	Model & (rad) & (cycle) & (rad) & (rad) & (cycle) & (rad) &  & (deg) & (deg) & (deg) & (deg) & ($d/R$) & \\
	\hline 
	Median     & 1.27  & -0.222 & 0.115 & 1.89  & -0.299 & 0.115 & 1.00 & 87.57 & 72 & 89 &  92 & 0.22 & [1] \\
	ML     & 1.28  & -0.255 & 0.151 & 1.70  & -0.323 & 0.124 & 0.82 & 87.71 & 78 & 95 &  69 & 0.22 & [1] \\
	Median & 1.573 &  0     & 0.101 & 1.578 &  0.553 & 0.100 & 1.00 & 87.61 & 90 & 90 &  92 & 0.17 & [2] \\
	ML     & 1.387 &  0     & 0.092 & 1.980 &  0.428 & 0.112 & 1.22 & 88.18 & 72 & 82 & 115 & 0.23 & [2] \\
	\hline
\end{tabular}
\tablefoot{The subscripts ${\rm p}$ and ${\rm s}$ stand for primary and secondary hot spot. See \cite{petri_polarized_2017} for the definition of the angles $(\alpha,\beta,\delta)$ and the displacement $\epsilon$.}
\end{table*}

\subsection{X-ray and $\gamma$-ray time lag}

Our light-curve fitting procedure relies heavily on multi-wavelength phase aligned pulse profiles. The most important characteristic of these profiles are the phase of the peaks in $\gamma$-ray and X-ray. For the radio pulse profile, the situation is less obvious, should we use the peak value or the centre of the radio pulse profile defined for instance by the interval containing at least $5\%$ or $10\%$ of the peak intensity. We think that the latter is more robust and choose the phase of the centre to lie halfway between the phases of $10\%$ peak intensity of the corresponding pulse.

Table~\ref{tab:phase_lag} summarises the locations of the pulse centres as defined above. Both X-ray pulse peaks arrive slightly before the middle of the radio pulse profile. This ordering is incompatible with the off-centred dipole geometry because the order of appearance of radio and X-rays should be inverted between north and south pole for simple circular polar cap shapes. 
This one-sided phase lag between X-ray and radio pulses requires an accurate quantitative geometric study of the location of the peak X-ray emission within the hot spot. On top of the geometry, to estimate the time lag, a careful analysis of aberration, retardation (A/R) and magnetic sweep-back effects must be included in a detailed analysis of the different wavelengths \citep{phillips_radio_1992}. This is however out of the scope of this work but see the discussion in Sec.~\ref{sec:discussion} for orders of magnitude estimates.

The X-ray pulse peaks are separated by approximately $0.4$ in phase. Looking at Table~\ref{tab:NICER1}, we therefore associate the first X-ray pulse denoted by P1X to the primary hot spot location at $\Theta_{\rm p}$ thus in the northern hemisphere, whereas the second X-ray pulse denoted by P2X is attributed to the secondary hot spot location at $\Theta_{\rm s}$ thus in the southern hemisphere. At radio frequencies, the peak separation is also about $0.45$ in phase, thus the second radio pulse denoted by P2R is linked to P1X whereas the first radio pulse denoted by P1R is linked to P2X. One X-ray pulse centre is leading its radio pulse centre by a phase shift about $0.1$, whereas the other X-ray pulse centre is almost aligned with its radio pulse centre.

\subsection{Radio emission and polarisation}

\begin{figure}[h]
	\centering
	\includegraphics[width=0.9\linewidth]{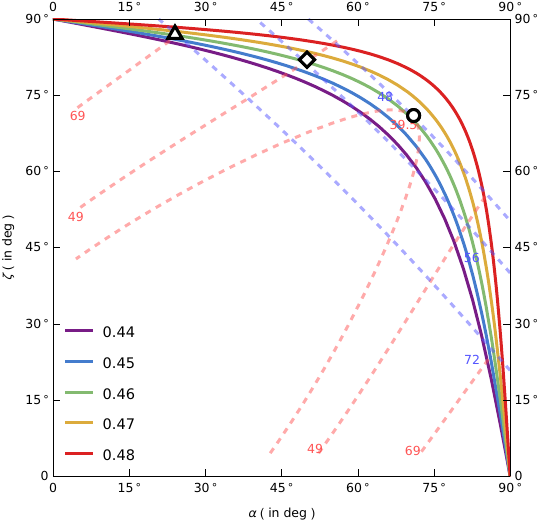}
	\caption{Constraints on the geometry with fixed radio pulse profiles. The red dashed curve constraints the width at 5\% maximum intensity given by $w_{5} = 0.233$ whereas the blue dashed curve constraint $w_{5} = 0.164$ for three values of $\zeta$.}
	\label{fig:anglecone}
\end{figure}

Some more information can be gained from studying the radio pulse profiles and their polarisation. It is well known that the width~$W$ is related to the beam cone half-opening angle $\rho$ by \citep{gil_geometry_1984}
\begin{equation}\label{key}
	\cos \rho = \cos \alpha \cos \zeta + \sin \alpha \sin \zeta \cos \, (W/2) \ .
\end{equation}
Moreover, if we assume that the full open field line region is producing radio photons then $\rho$ is related to the polar cap size by $\rho \approx \sqrt{R/\rlight}$. The pulse widths at 5\% maximum intensity for the first and second radio peak are respectively $w_{5} = \{0.164, 0.233\}$. The pulse width imposes some constraints on $\alpha$ and $\zeta$ as shown in Fig.~\ref{fig:anglecone}. In order to estimate the beam opening angle, we compare the three geometries, namely minimalistic $\zeta=\alpha = 71 \degr$, best $\gamma$-ray fit $(\alpha,\zeta)=(51\degr,82\degr)$ and NICER fit $(\alpha,\zeta)=(24\degr,87\degr)$. The corresponding radio beam cone half-opening angle for one pole is then respectively $39.5\degr, 49\degr, 69\degr$ and for the other $48\degr, 56\degr, 72\degr$. 
For these inclination angles we get for the P2 width $w_{5} = 0.233$ opening angles as $39.5\degr, 49\degr, 69\degr$ whereas for the P1 width $w_{5} = 0.164$ opening angles as $48\degr, 56\degr, 72\degr$. 
Values are summarised in Table~\ref{tab:cone_opening_angle}. These angle can be related to emission heights~$h_{\rm e}$ \citep{petri_multi-wavelength_2024-3} with
\begin{equation}\label{eq:emission_height}
	\cos \rho = \frac{2-3\,h_{\rm e}/\rlight}{\sqrt{4-3\,h_{\rm e}/\rlight}},
\end{equation}
and also reported in the same table. The emission heights correspond to a substantial fraction of the light-cylinder radius, between 20\% and 50\% of $\rlight$. At these altitudes, some plasma effects will distort the dipole field and the approximation used in Eq.\eqref{eq:emission_height} becomes inaccurate.
\begin{table}[h]
\centering
\caption{Radio beam cone opening angle $\rho$ according to different geometries of the pulsar giving $\rho$ in the first part and $h_{\rm e}/\rlight$ in the second part separated by a '/'.\label{tab:cone_opening_angle}}
\begin{tabular}{cccc}
	\hline
	Pulse & Min & \textit{Fermi} LAT & NICER \\
	 & (71\degr,71\degr) &  (51\degr,82\degr)&  (24\degr,87\degr)\\
	\hline	
	P1 & 48\degr / 0.27 & 56\degr / 0.35 & 72\degr / 0.50 \\
	P2 & 40\degr / 0.19 & 49\degr / 0.28 & 69\degr / 0.47 \\
	\hline
\end{tabular}
\end{table}

\begin{figure}[h]
	\centering
	\includegraphics[width=0.8\linewidth]{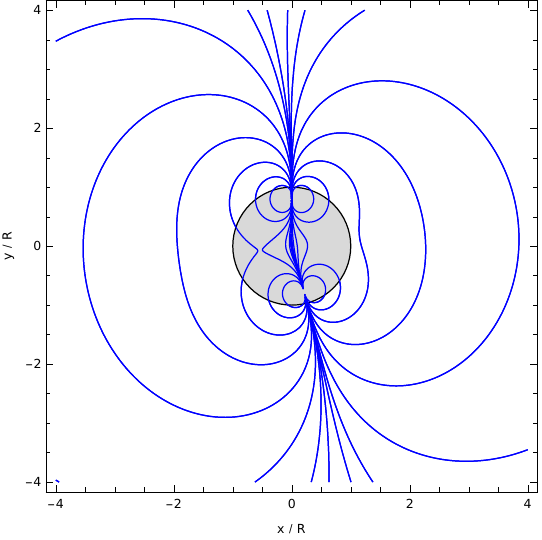}
	\caption{An example of double dipole geometry in the equatorial plane. Magnetic field lines are shown in blue and the neutron star surface as a solid black circle of normalised radius. The stellar interior is shown in light gray.}
	\label{fig:doubledipole}
\end{figure}

Moreover, the cone opening angle~$\rho$ is not compatible with the hot spot angular sizes~$\xi_{\rm p}$ and $\xi_{\rm s}$ found by the NICER collaboration. Indeed if we assume that the rim of the almost circular hot spot corresponds to the edge of the radio emission beam then the emission height~$h$ is given by 
\begin{equation}
	\frac{h}{R} =  \frac{\sin^2 \theta}{\sin^2 \xi_{\rm p,s}}
\end{equation}
with $\theta$ the colatitude at which the radio beam has an opening angle $\rho$. For a pure static dipole, both angles are related by
\begin{equation}
	\rho = \theta + \arctan (\tan \theta/2) \ .
\end{equation}
For $\xi_{\rm p,s} \lesssim0.125$, we always find $h\gtrsim \rlight$, the radio emission would be produced outside the light cylinder. One possible conclusion is that the hot spot radiating X-ray are not directly connected to the large scale dipole field, because a bombardment of particles onto the polar caps would lead to much larger circular spots of angular size $\xi \sim 0.3$. This also means that the magnetic moment vector does not necessarily intersect the hot spots and therefore the magnetic obliquities given in Table~\ref{tab:NICER1} are subject to caution. Another possibility is to split the dipole into a double dipole geometry where two dipoles coexist and are located just underneath the surface of both hot spots. Figure~\ref{fig:doubledipole} shows an example of a double dipole geometry contained in the equatorial plane with two dipoles of same magnitude. The field lines look very similar to a single dipole at large distance but close to the surface, the non antipodal location of the magnetic pole is visible. We explore this idea deeper in the discussion section below.

\begin{figure}[h]
	\centering
	\includegraphics[width=\linewidth]{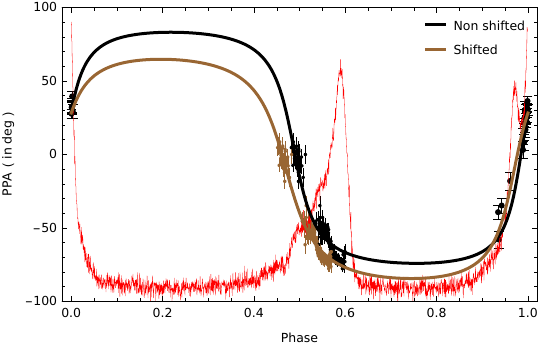}
	\caption{PPA as a function of rotational phase for PSR~J0740+6620 as measured with the NRT at 1.4~GHz. The best values for the non shifted PPA are $\alpha=79\degr$ and $\zeta=89\degr$ whereas for the PPA shifted in phase by $0.03$ it is $\alpha=74\degr$ and $\zeta=88\degr$. The radio pulse profile is shown in red for better identification of both pulses.}
	\label{fig:polarisation}
\end{figure}

The PPA evolution can help to check the consistency of the magnetic geometry. In Fig.~\ref{fig:polarisation} we show PPA values as obtained from a high signal-to-noise ratio observation of PSR~J0740+6620 conducted with the NRT at 1.4~GHz, on MJD~59576 (28 Dec 2021). The Stokes parameters were calibrated using the method presented in \cite{guillemot_improving_2023}, and were corrected for Faraday Rotation, using the Rotation Measure (RM) value of $-31$~rad~m$^{-2}$, determined from the data themselves using the \texttt{rmfit} tool of PSRCHIVE \citep{hotan_psrchive_2004}. Relying on the rotating vector model of \cite{radhakrishnan_magnetic_1969}, Fig.~\ref{fig:polarisation} shows a good fit to the data with $\alpha=79\degr$ and $\zeta=89\degr$, fitting simultaneously P1R and P2R. We assumed that the PPA shows orthogonal mode switching for P2R depending on the phase, one mode with $0.4<\phi<0.53$ and the other with $0.53<\phi<0.6$. Because P2R is not shifted by half a period compared to P1R, we also tried a fit by translating P2R to earlier phases by an amount of $\Delta\phi=-0.03$ (brown shifted curve). This ensures the antipodal position required for the RVM fitting formula instead of fitting P1R and P2R separately and individually. In the shifted approach, one single set of parameters is sufficient for both pulses. The associated fit is slightly better than the unshifted case but the geometry remains almost unchanged with $\alpha=74\degr$ and $\zeta=88\degr$ instead of $\alpha=79\degr$ and $\zeta=89\degr$ for the unshifted black curve.

\section{Discussion\label{sec:discussion}}

The results from our $\gamma$-ray light-curve fitting and from the NICER results are subject to some caveats. First, the \textit{Fermi} LAT peak separation relies on the phase difference between the peak intensity of both pulses from which $\Delta$ is deduced. However, for PSR~J0740+6620, the profile is asymmetric between the leading and trailing edge, rising sharply and decreasing much smoother in both peaks. This raises the question about the precise location of the centre of each peak. Should the peak separation $\Delta$ be computed according to the peak intensity as done in Fermi 3PC, or should it be extracted from the phase centre of each pulse defined by a width at 10\% maximum intensity for instance? 
If we allow for an uncertainty in the estimate of $\Delta$, from the plot in Fig.~\ref{fig:obliqinclin}, we stress that at large line of sight inclination angle $\zeta \gtrsim80 \degr$, the obliquity $\alpha$ becomes very sensitive to the value of $\Delta$. A slight increase in $\Delta$ can lead to a drastic increase in $\alpha$. For instance choosing $\Delta=0.48$ and keeping $\zeta=82\degr$ we would find $\alpha=66 \degr$ much closer to the $72\degr$ found by NICER.

Secondly, the NICER collaboration assumes that the orbital plane inclination angle $i$ is equal to the observer line of sight inclination angle $\zeta$, thus $i=\zeta$. But this is valid only if the pulsar had enough time to exactly align its spin axis with the orbital angular momentum. Whether the accretion phase was efficient and long enough to proceed to exact alignment is unclear, a small deviation of several degrees might be possible. Third, the magnetic axis was assumed to cross the centre of each hot spot, implying implicitly that the heating of the polar caps by particle bombardment is circular symmetric. However, this assumption too needs to be corrected as there should be a sweep back of magnetic field lines and some other effects related to the stellar rotation. Therefore the values for $(\alpha, \beta, \delta, \epsilon)$ shown in Table~\ref{tab:NICER1} should be reevaluated in light of these caveats.

Finally, the angular size of both hot spots are very similar $\xi_{\rm p} \approx \xi_{\rm s} \approx 0.115$. However the usual size of the polar cap, deduced from the opening field line region gives a value of $\theta_{\rm pc} \approx \sqrt{R/\rlight} \approx 0.306$ thus almost a factor three larger for this estimate about a centred dipole. As we showed in \cite{petri_constraining_2023-2}, an off-centring of the dipole location can drastically reduce the size of the polar cap. If for simplicity we assume an aligned rotator and shift the dipole along the rotation axis to the south pole, the displacement $z_0$ required to adjust the spot size to an angle $\xi_{\rm p}$, and dropping the negative solution, is given by
\begin{equation}
	z_0 / \rlight = - \cos \xi_{\rm p} \sin^2 \xi_{\rm p} + \sqrt{(R/\rlight)^2 - \sin^6 \xi_{\rm p}} \ .
\end{equation}
Putting numbers adapted for PSR~J0740+6620, we find that a shift of $z_0 / \rlight \approx 0.077$ is sufficient to reduce the polar cap size to the observed value. This corresponds to a distance $z_0 \approx 10.7-11.1$~km, thus to a size related to the neutron star radius of $z_0/R \approx 0.85-0.89$. The dipole is located slightly underneath the surface, in the crust. The same idea can be applied to the south polar cap with a similar shift.

Because the light-cylinder radius is only $11$ times larger than the stellar radius with $R/\rlight \approx 0.091$, the impact of an off-centred or a double dipole remains substantial for the geometry of the current sheet in the striped wind. Indeed the perfect symmetry in the light-curves is broken for those configuration and the relation between the angles $\alpha$ and $\zeta$ and the peak separation $\Delta$ deviates slightly from the analytical expectations given by Eq.\eqref{eq:cos_pi_delta}. Variation in $\Delta$ of about $0.01-0.02$ are easily achievable for MSPs. The impact on the obliquity~$\alpha$ for high inclination angles~$\zeta$ around $85 \degr$ is dramatic as shown in Fig.~\ref{fig:obliqinclin}. Setting for instance the value for $\zeta$ approximately to $82\degr$ or $87\degr$ we find the obliquities shown in Table~\ref{tab:Geometrie}. The peak separation is not the most reliable property of the light-curves, especially if the leading and trailing edges of the pulses are asymmetric. The scrutiny of the full shape is therefore recommended because the pulse centre then shifts from the peak intensity. Nevertheless, within a few degrees of uncertainties, the radio emission and PPA fit agrees with the $\gamma$-ray light curve and with the location of the thermal hot spots. This gives us confidence about the robustness of the double dipole approach for such a recycled pulsar.
\begin{table}[h]
	\centering
	\caption{Obliquity $\alpha$ of the magnetic field for varying peak separation~$\Delta$ and different inclination angle $\zeta$.}
	\label{tab:Geometrie}
	\begin{tabular}{cccccc}
		\hline 
		$\Delta$ & $0.46$ & $0.47$ & $0.48$ & $0.49$ & $0.5$\\
		\hline
		$\zeta = 82\degr$ & $48\degr$ & $56\degr$ & $66\degr$ & $77\degr$ & $90\degr$ \\
		$\zeta = 87\degr$ & $23\degr$ & $29\degr$ & $40\degr$ & $59\degr$ & $90\degr$ \\
		\hline
	\end{tabular}
\end{table}

\begin{figure}
	\centering
	\includegraphics[width=0.95\linewidth]{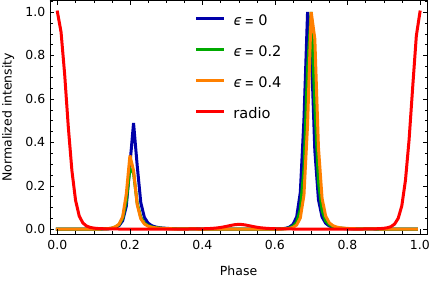}
	\caption{Comparison of the simulated $\gamma$-ray light curves between the centred dipole in blue and the off-centred dipole in green and orange for $\alpha=70\degr$, $\beta=0\degr$, $\delta=90\degr$ and $\epsilon=0.2$. The line of sight inclination is $\zeta=82\degr$. The radio pulse profile is shown in red.}
	\label{fig:gamma_decentre}
\end{figure}

To be more quantitative, Fig.~\ref{fig:gamma_decentre} shows an example of variation of the $\gamma$-ray peak separation~$\Delta$ when an off-centred dipole is taken into account. We computed the force-free magnetosphere solutions for $\alpha=70\degr$, $\beta=0\degr$, $\delta=90\degr$ and $\epsilon=\{0.2,0.4\}$ and then extracted the $\gamma$-ray light-curves as presented in Fig.~\ref{fig:gamma_decentre}. In this particular case with $\zeta=82\degr$, the off-centring has increased the peak separation~$\Delta$ by approximately $0.02$ in phase for both eccentricities ($\epsilon=0.2$ and $\epsilon=0.4$) compared to the centre dipole, as shown by the green and orange curves that almost overlap with this viewing angle. Therefore for fast rotating MSPs, the peak separation measurements are not sufficient to robustly estimate the geometry of the large scale dipolar magnetic field. Some cross-check by using multi-wavelength data is required as we demonstrated in this work.

Finally, knowing the radio emission height denoted by $r_2$ and setting the thermal X-ray emission radius to $r_1=R$, we can estimate the delays introduced by aberration $\Delta t_{\rm a}$, retardation $\Delta t_{\rm r}$ and sweepback $\Delta t_{\text{\tiny B}}$ effects for which the magnitude increases with increasing altitude according to \cite{phillips_radio_1992} because
\begin{subequations}\label{key}
	\begin{align}
	\frac{\Delta t_{\rm r}}{P} & = \frac{r_1-r_2}{2\,\pi\,\rlight} \\
	\frac{\Delta t_{\rm a}}{P} & = \frac{\arctan \left[(r_1/ \rlight ) \sin\rchi \right] - \arctan\left[ ( r_2 / \rlight ) \sin\rchi \right]}{2\,\pi} \\
	\frac{\Delta t_{\text{\tiny B}}}{P} & = \frac{1.2\,\sin^2\rchi}{2\,\pi} \, \left[\left( \frac{r_2}{\rlight} \right)^3 - \left( \frac{r_1}{\rlight} \right)^3 \right] \ .
	\end{align}
\end{subequations}	
From the previous section, we found a maximum radio emission height about $40-50$\% of $\rlight$ at which aberration and retardation effects contribute approximately equally to $6$\% of phase advance each, whereas vacuum magnetic field sweepback contributes approximately $2.5$\% of phase retardation leading to a total of $10$\% phase advance of the radio pulse compared to the thermal X-ray emission. If however we use the more realistic force-free magnetic field sweepback effect instead of the vacuum one, the term involving $\Delta t_{\text{\tiny B}}$ contributes approximately $6$ times more than in vacuum (see for instance figure~6 of \cite{petri_multi-wavelength_2024-3}) thus adding $15$\% to the phase retardation. In all in all, the force-free sweepback compensates the A/R phase advance leading to almost phase-aligned radio and X-ray as observed for PSR~J0740$+$6620. More detailed calculations would require an exact determination of the radio emission height and the full force-free magnetosphere geometry using for instance the double dipole configuration shown in figure~\ref{fig:doubledipole}. However, we do not dive into such refinements for the present study.

\section{Conclusions\label{sec:conclusion}}

Although MSP emission is usually thought to be difficult to model with a dipolar magnetic field because of the impact of multipolar components close to the stellar surface, we showed that a double dipole geometry can confidently reproduce a wealth of multi-wavelength properties of PSR~J0740+6620. A mildly off-centred dipole would be compatible with radio, thermal X-ray and $\gamma$-ray pulse profiles but the size of the heated polar cap would be three times larger than the expected size deduced from the NICER collaboration. Both dipoles are almost antipodal and located approximately $10\%$ below the surface. Whether such a configuration could be applied to other MSPs must be tested case by case. Good candidates are those from other NICER pulsars published results like PSR~J0437-4715 \citep{choudhury_nicer_2024} or PSR~J1231-1411 \citep{salmi_nicer_2024}. We plan to investigate these pulsars in a near future.

\begin{acknowledgements}
We are grateful to the referee for helpful comments and suggestions. 
This work has been supported by the ANR (Agence Nationale de la Recherche) grant number ANR-20-CE31-0010. SG and DGC acknowledge the support of the CNES.
This paper is partially based on data obtained using the NenuFAR radio-telescope. The development of NenuFAR has been supported by personnel and funding from: Observatoire Radioastronomique de Nan\c{c}ay, CNRS-INSU, Observatoire de Paris-PSL, Université d’Orléans, Observatoire des Sciences de l’Univers en Région Centre, Région Centre-Val de Loire, DIM-ACAV and DIM-ACAV + of Région Ile-de-France, Agence Nationale de la Recherche.
The Nan\c{c}ay Radio Observatory (NRT) is operated by the Paris Observatory, associated with the French Centre National de la Recherche Scientifique (CNRS) and Université d’Orléans. It is partially supported by the Region Centre Val de Loire in France.
We acknowledge the use of the Nan\c{c}ay Data Centre computing facility (CDN -- Centre de Données de Nan\c{c}ay). The CDN is hosted by the Observatoire Radioastronomique de Nan\c{c}ay (ORN) in partnership with Observatoire de Paris, Université d'Orléans, OSUC, and the CNRS. The CDN is supported by the Région Centre-Val de Loire, département du Cher.
The Nan\c{c}ay Radio Observatory (ORN) is operated by Paris Observatory, associated with the French Centre National de la Recherche Scientifique (CNRS) and Universit\'{e} d'Orl\'{e}ans.
\end{acknowledgements}


\end{document}